# *Thi*ng/*Mac*hine-*s* (*Thimacs*) Applied to Structural Description in Software Engineering

Sabah Al-Fedaghi
Computer Engineering Department
Kuwait University
Kuwait
sabah.alfedaghi@ku.edu.kw

*Abstract*—We are pursuing a modeling methodology that views the world as a realm of *things*. A thing is defined as something that can be created, processed, released, transferred, and received. Additionally, in this modeling approach, a thing is a five-dimensional structure referred to as a thinging (abstract) *machine*. On the other hand, machines are things that are "operated on"; that is, they are created, processed, released, transferred, and received. The intertwining with the world is accomplished by integrating these two modes of an entity's being: being a thing that flows through machines and being a machine that processes things. This paper further enriches these notions of things and machines. We present further exploration of the thinging machine model through introducing a new notion called the *thi*ng/*mac*hine (*thimac*) as a label of the unity of things/machines. Thimacs replace traditional categorization, properties, and behavior with creating, processing, releasing, transferring, and receiving, as well as the two linking notions of flow and triggering. The paper discusses the concept of thimacs with examples and focuses on the notion of structure as it applies to various diagrammatic modeling methodologies.

## I. INTRODUCTION

Conceptual modeling involves identifying, analyzing, and describing the essential concepts and constraints of a domain with the help of a (diagrammatic) modeling language [1]. The term *conceptual* refers to *conceptualization* as "an abstract model of the things that are assumed to exist in some area" [2]. A diagrammatic language is adopted on the assumption that in software engineering, "it is almost impossible to model without a conceptual diagram to visualize the modeler's concepts and the system" [3]. Diagram-based conceptual modeling typically embed the notions of input, output, and process, which form the traditional input-process-output model used in many interdisciplinary applications.

A model in this context involves observing the "world," determining what is relevant, choosing/defining terminology, writing down the governing laws, and verifying the correctness/completeness of these laws [4]. Modeling is the foundation of all engineering disciplines, as it improves communication and understanding about why the system is needed, what its functionality should be, and how it should be implemented [5].

Research on modeling "real world" features leads modelers to philosophy, the "mother of all sciences" [6]. As Hassan et al. [7] notes, understanding philosophical questions can help to ensure that the research is rigorous and insightful. "All research is philosophy in action" [7].

In this paper, we study modeling using the Thinging Machine (TM) model, which has been used in a series of research papers [8-19].

### A. Further Exploration of the Philosophy of the Thinging Model

In software engineering, modeling activities involve building either software models (requirements specification) or "real world" (domain) models [20]. This latter form of modeling is the focus of this paper, with an emphasis on a domain model that represents the real world and is used later as a specification of the system to be constructed (software model) or as a tool for simulation. Additionally, "understanding the real world most useful to produce a good user requirements specification, and therefore a practical software system that solves the needs of users" [20].

According to Gruner [21], the achievements of software engineering made it possible to induce a global social evolution. Nevertheless, partly due to this success and also due to the problems, shortcomings, and failures experienced, a need for philosophical theory of software science and engineering began to emerge [21]. As Pyshkin notes [22], "Despite many formalized ways to represent data models, program structures, execution analysis and verification, and project organization were discovered, software engineering is still far from being an exact science." "Up to now, a concise, practicable theory of software engineering does not exist'' [23], and its philosophical foundations and premises are not yet well understood. The philosophy of software engineering is related to the philosophy of computer science, but they have their differences; they "are not identical branches of a general philosophy of science" [21].

In this context, O'Bower [24] claims that a software engineer [programmer] is not a mathematician but rather a philosopher and linguist all in one ([24], as reported in [22]). Arkhipenkov [25] declares that "software development is a kind of human activity which is mistakenly attributed to engineering" [25].

According to Ventura [26], "we should ask whether there is a philosophical foundation for software design principles…

What if Aristotle Were a Software Engineer?" Imagine that Aristotle is living in our time, working as a software engineer and dealing with common problems that emerge during the process of software design. Computer science involves the study of computational *artifacts*, the artifacts concerned with the nature, specification, design, and construction of a system. The abstract nature of the implemented artifacts ensures that many of the conceptual questions have analogues in philosophy [27].

Butcher [28] explored the philosophical systems that Plato and Aristotle devised, showing their relations to object-oriented programming (OOP). It seems that if we examine all major OOP ideas, their philosophical origins will eventually be found [26]. Interest in OOP has dominated research because it is claimed that OOP is "more natural" due to its similarity to Aristotelian logic [29].

In the classical description of the object-oriented model, the real world is viewed as being made up of objects; therefore, object orientation (OO) is the most natural way in which to understand the world and to develop software systems that solve human needs [30]. This object-oriented description seems to be derived from classic philosophical ideas. For example, the central notion in Aristotelian philosophy is an entity: the living being that cannot be decomposed into parts. The danger appears when it is critically accepted that "everything is an object" [20]. Génova [20] proclaims:

> The notion of object in software is inspired rather on the mechanical device, an assembly. Operations are performed on the objects, whilst in the Aristotelian world; operations are performed by the entities. Moreover, there are other competing philosophical views of the world, where the central notion is not the object or entity, but rather the process… If we only want to build information systems, all we need to understand is *the world as we speak about it*, the information world.

The achievement of the OO approach in software engineering is associated with the proliferation of OO technology. According to Duckham [31], "this proliferation has not always been complemented by a growth in OO theory. The surfeit of object-oriented analysis, design and programming techniques which exist are, therefore, necessarily highly subjective." According to Joque [32]:

> Despite the obvious allusion to object-oriented programming in the naming of object-oriented ontology, there are few descriptions of the relationship between object-oriented programming and said ontology. This is especially unfortunate as the history and philosophy that surround object-oriented programming offer a nuanced understanding of objects, their ability to hide part of themselves from the world, their relations, and their representation in languages that in many ways challenge the claims offered by object-oriented ontology.

Although such issues related to the philosophy of software engineering are still controversial, they motivate further exploration of alternative fundamental modeling approaches. The diversity of basic modeling approaches should benefit conceptual modeling research and could also advance current proven practices, such as OO.

Specifically, this paper further explores the expressive power of the TM as a modeling language for use in the difficult philosophical endeavor of describing systems. The paper's aim is to contribute to previous claims in this line of research regarding the capabilities of the TM as a modeling tool in diverse domains.

The TM relies more on Heidegger's [33] notion of *things* than it does on objects. Heidegger's works on thinging have been applied in various scientific fields (e.g., design thinking [34] and information services [35]). In this paper, we analyze TM diagramming and provide several examples that have been re-modeled in terms of the TM to reveal a rich conception of the various aspects of the TM model.

*B. New Contribution: The Thimac Integrates Things and Processes*

In this paper, we follow the TM methodology, which views the world as a world of things. As a sequel of a previous paper [8] about the TM describing behavior in a system, this paper emphasizes the *structural description* of the modeled system. The main thesis is that each entity has a double nature as (i) a thing and (ii) a process (abstract machine); thus, we call thing-machines (*thimacs*). In modeling, the intertwining with the world is accomplished by integrating these two modes of being of an entity: being a thing that flows through machines and being a machine that processes things. Thimacs inhibit the application of traditional categorization, properties, and behavior, replacing them with creating, processing, releasing, transferring, and receiving, as well as the two linking notions of flow and triggering.

We trace this idea to Aristotle in history and to Heidegger in modern times. We then apply their ideas to formulate an evolving interpretation of the TM. We re-model several structural representations from the literature using the TM concept. In general, our approach in this paper is to place the various diagrammatic representations of systems side by side to assist in understanding the various features of the models.

Accordingly, the next section further enriches the notions of things and machines presented previously in Al-Fedaghi [1-19] and enhances them with examples. Section 3 discusses the new concept of thimacs, with an example given in Section 4. The rest of the paper focuses on the notion of structure as it applies to the UML composite structure diagram and other constructs.

II. THINGS AND MACHINES

We start our voyage about the notion of things with Aristotle. Things in the world captivated him, as he famously said, "In all things of nature there is something of the marvelous." Aristotle said that it "behooves us to begin philosophizing by laying out the *phainomena*, the *appearances*, or, more fully, *the things appearing to be the case*" [36]. His method is to reflect upon the *aporiai* of a thing (e.g., time): If the thing exists, then what sort of thing is it? Is it the sort of thing to exist absolutely and independently? Or, is it the sort of thing that depends upon other things for its existence [36]?

Aristotle proclaimed that things of the sort *man* and *runs* signify entities that correspond to linguistic structures such as the sentence *Man runs*. The entities are the sorts of "basic beings that fall below the level of truth-makers, or facts, just as,…nouns and verbs, things said 'without combination', contribute to the truth-evaluability of simple assertions" [36]. Additionally, he introduced the notion of *process* in thinking about things. He conjectured that a thing in nature persists via an internal process that must be realized within a matter that harbors tendencies resulting from its elemental components (e.g., fire, water, earth, or air). This causes tendencies to actively strive toward their "natural place." In this view, Aristotle can be counted as a process philosopher [37]. Aristotle's idea is that things are compounds consisting of *matter* and *form* (e.g., a statue is a compound of bronze and its perceptible shape).

Due to space considerations, we abruptly close this discussion with Heidegger [33]. For Heidegger, things have unique "thingy Qualities" that are related to reality but are therefore not typically found in *objects*. According to Heidegger [33], a thing is self-sustained, self-supporting, or independent—something that stands on its own. The condition of being self-supporting transpires by means of *producing* the thing.

Heidegger [33] encourages further research on "generic processes" applied to a thing. We claim that five processes of things exist: They can be created, processed, released, transferred, and received. For instance, suppose that *t* is a thing. To describe the elemental processes [38] that can be applied to *t* in a given system, *S*, the following discussion presents an informal justification for them.

- Thing *t* either comes from the outside of *S* (**transferred in**) or is generated internally (**created**).
- When *t* is transferred from outside of *S*, it is either rejected or **received** as one of the system's things.
- Thing *t* in *S* may be **transferred outside** of *S*.
- The thing may be put in the **released** state until a channel is open for transfer it to the outside.
- During its residency in *S*, *t* may be **processed** (changed).

These elementary processes have been called various names. For example, the *Merriam-Webster Dictionary* (https://www.merriam-webster.com/thesaurus/create) lists the following synonyms for the word *create*: beget, breed, bring, bring about, bring on, catalyze, cause, do, draw on, effect, effectuate, engender, generate, induce, invoke, make, occasion, produce, prompt, result (in), spawn, translate (into), work, and yield. However, several of these words do not mean *create* in the TM sense. Consider *work* as in *John works*. In the TM, this word indicates that John *creates* and *processes* works. *Process* here means that the created work takes its course. Consider the clause "The engine works"; in the TM; the engine has a *state*, and the state *work* is created (instead of the state *not work*).

These processes have been applied successfully to many real systems, such as phone communication [15], physical security [12], vehicle tracking [14], computational thinking [9], business processes [10], and intelligent monitoring [11].

Thus, we assert that we can use these five generic processes to model things. We also define a thing in terms of these five processes: created, processed, released, transferred, and received.

Additionally, a thing flows in an abstract, five-dimensional structure that can be referred to as the TM. These elementary processes form a complex process (abstract machine) called a TM as shown in Fig. 1, where the elementary processes are called the *stages* of a TM. A TM can be put into the form of the input–process–output model (Fig. 2).

Flow (solid arrow) among the five stages in Fig. 1 signifies *conceptual* movement from one machine to another or among the stages of a machine. The TM stages can be described as follows:

**Arrived**: A thing reaches a new machine.
**Accepted**: A thing is permitted to enter the machine. If arriving things are always accepted, *Arrive* and *Accept* can be combined as the **received** stage. For the purpose of simplification, the examples in this paper assume the received stage.
**Processed** (changed): A thing undergoes some kind of transformation that changes it without creating a new thing.
**Released**: A thing is marked as ready to be transferred outside of the machine.
**Transferred**: A thing is transported somewhere from/to outside of the machine.
**Created**: A new thing is born (created) in a machine in an analogy to the dynamic creation of objects in the UML. The term *create* comes from creativity with respect to a system (i.e., constructed things from already created things, or emergent things appear from somewhere.

Additionally, the TM model includes *memory* and **triggering** (represented as dashed arrows) relations among the processes' stages (machines).

The following examples illustrate the expressive power of the TM language.

**Example 1:** According to Aristotle, "For a thing to have a nature is for it to have an inner source of changing and of staying the same. An oak tree, for instance, has a nature; a bed does not" [39].

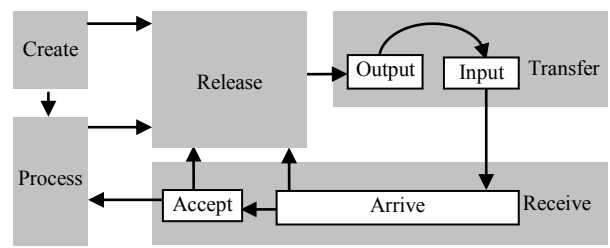

Fig. 1. Thinging machine.

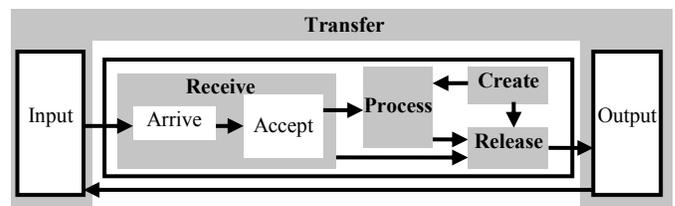

Fig. 2. Another description of a TM.

We view Aristotle's nature as a TM of a special kind related to its *create* ability. *Create* in the TM is a loaded notion. For example, it may donate the mere appearance by, for instance, creating a record from data. *Create* may also denote becoming; that is, the machine produces a completely new thing, much as a man creates concepts. The last sense falls is Aristotelian in nature.

> The oak tree has certain characteristic ways of changing: It loses its leaves in autumn, it grows acorns, it puts out roots of a certain shape. According to Aristotle, these characteristic changes are…by the stuff of which the tree is made. [39]

As shown in the TM representation of Fig. 3, the tree has inner changes: creating leaves, roots, etc. Fig. 3 embeds a structure in the sense that a whole/part relationship exists. It also embeds "behavior" given that some things flow. The time of autumn does not yet appear in the picture. This is discussed when we introduce the notion of events.

**Example 2:** According to Coope [39], "In contrast, a bed does not have characteristic ways of changing. There are no changes that it undergoes of itself in virtue of being a bed. Its changes are caused partly by its environment and partly by the stuff of which it is made. If we bury a bed and its rotting frame puts out shoots, this will be because it is wooden, not because it is an ex-bed. The bed will not spawn baby beds."

As shown in the TM language in Fig. 4, the bed generates changes that stem partly from its environment and partly from the stuff of which it is made.

**Example 3**: According to Aristotle, the potencies of things are "originative sources of change in another thing or in the thing itself qua other" [40]. Two kinds of potencies of things exist: the passive potency to be acted upon and the active potency to act on (a potency found in agents to impart change). Inanimate things possess only the first kind; organisms, on the other hand, have both.

However, "in so far as a thing is an organic unity, it cannot be acted on by itself; for it is one and not two different things" (see [40], p. 17). Because nothing can possess both potencies, nothing can act on itself. "The whole moves itself not by virtue of having some part such as to move itself; it moves itself as a whole, moving and being moved by virtue of part of it moving and part of it being moved. It does not move as a whole, and it is not moved as a whole" [40].

Consider an animal's self-motion, such as a lion pursuing a gazelle. To realize the movement, the soul reaches out toward the lion and actualizes the (passive) potency of the body. The active aspect of the lion (its soul) efficiently causes the passive aspect (its body) to move. The lion's psyche is the unmoved mover of its body. The lion's soul thus requires an external object (the gazelle), as intentionally represented, to actualize the soul's desire by serving as the object of desire. Aristotle therefore explains an animal's self-motion by splitting the organism into two: the soul (the unmoved mover) and the body (the moved).

Fig. 5 shows the TM model in this case. The lion has a body and spirit (circles 1 and 2, respectively). The sight of the gazelle (3) triggers the spirit (4), which prompts the movement (5) of the body to pursue the gazelle (6).

Thus, the Aristotelian principle that nothing can cause itself applies in this case: "Aristotle wanted to make sure the universe is not understood as bringing itself into existence" [40]. In Descartes's dualism, a nonphysical mind replaces the spirit, as the agent has efficient cause for activating the body. Materialism analyzes voluntary motion as the brain activates it. One part of the body triggers another, which pushes a third, and so on "until something shoves the skeletomuscular system into action" [40].

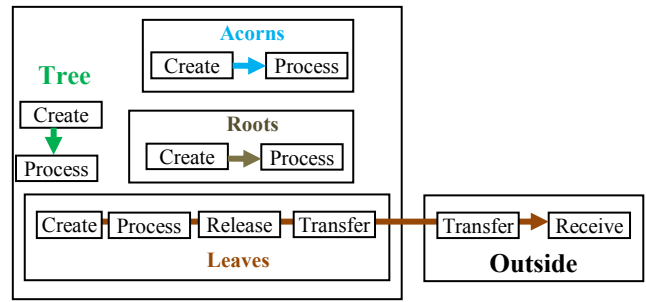

Fig. 3. The oak tree has certain characteristic ways of changing; it loses its leaves in autumn, it grows acorns, and it puts out roots of a certain shape.

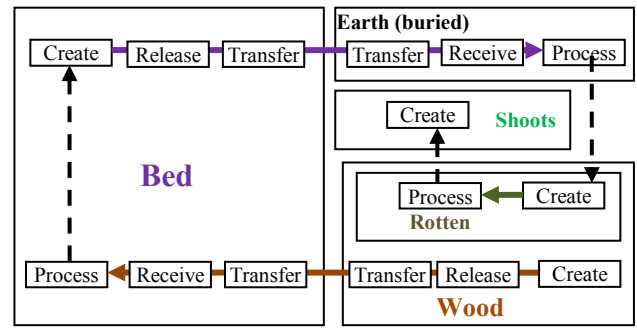

Fig. 4. If we bury a bed and its rotting frame sprouts new shoots, this is because it is wooden.

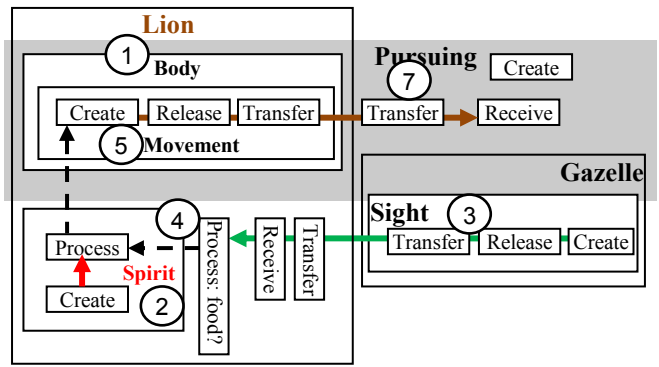

Fig. 5. A model of a lion's self-motion.

This example provides an opportunity to discuss the dynamism of the TM model; we examine the notion of *events* with a focus on the interaction of events with such concepts as time, space, and "regions" in the TM diagram. The notion of *events* has occupied a central role in modeling and has an influence on software engineering and philosophy.

An event plays a prominent role in various areas of philosophy, from metaphysics to the philosophy of action and the mind, as well as in such diverse disciplines as linguistics, literary theory, probability theory, artificial intelligence, physics, and—of course—history. This plethora of concerns and applications is indicative of the prima facie centrality of the notion of an event in our conceptual scheme. [41]

Whitehead declared that everything is an event. The world is made of events and nothing but events. Even a solid object is an event or, better, a multiplicity and a series of events (see [42]).

An event in a TM is modeled by its *region* (i.e., where the event occurs). *Time* and *itself* are exemplified by the event *A gazelle appears and is seen by the lion* (see Fig. 6). For simplicity's sake, we represent an event only by its region. Events form a hierarchy in which larger events include smaller events. For example, Fig. 7 shows four "meaningful" events in the model of a lion pursuing a gazelle.

Event 1 ($E_1$): A gazelle appears and is seen by the lion.
Event 2 ($E_2$): The lion realizes the gazelle's potential as food.
Event 3 ($E_3$): The lion's spirit is triggered.
Event 4 ($E_4$): The movement of pursuing is actualized.
Fig. 8 shows the chronology of these events.

### III. THE THING/MACHINE (THIMAC)

In the TM approach, a thing is not just an entity. It is also a machine that handles other things. Machines are also "cooked" (i.e., operated on); that is, they are created, processed, released, transferred, and received. In thins/machine, the intertwining with the world is accomplished through both of these modes of being: being an entity that flows through machines and being a machine through which other things flow.

A *thing/machine* is denoted here as thimac; *thi* stands for thing and *mac* for machine (see Fig. 9). The thimac is a five-dimensional *structure*: creating, processing, releasing, transferring, and receiving. Furthermore, the model system is a (grand) thimac forming the "space" of the system filled with *sub-thimacs*; space in this context is used in the analogy of points of space as locations of space.

A thimac is understood to be *structured* in the sense that various relations are defined on the sub-thimacs. It is these relations that one aims to characterize when one speaks of a "thimac structure." The thimacs "communicate/are tied together" through flows and triggering.

The Aristotelian *form*, mentioned previously, is a type of machine. For example, in the bronze statue, its thingness is formed by its machineness as shown in Fig. 10. The notion of structure can be applied to the thing and the machine. A thing can have a subthing as shown in Fig. 11, where the water *mac* forms its structured *thi* from oxygen and hydrogen.

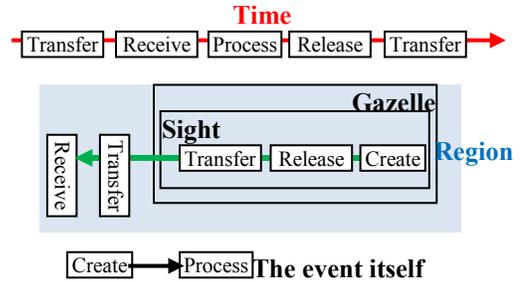

Fig. 6. The event *A gazelle appears and is seen by the lion*.

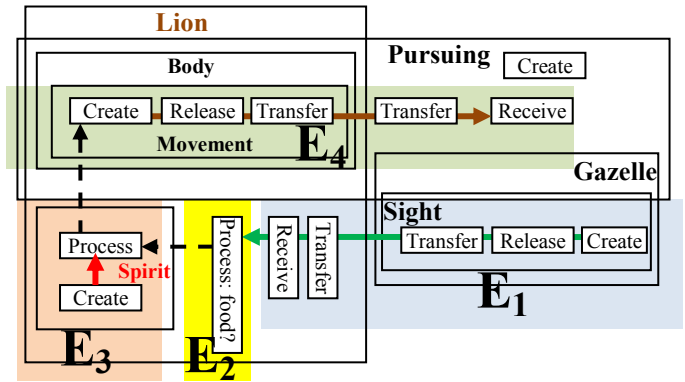

Fig. 7. Events of a lion pursuing a gazelle.

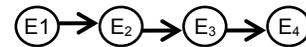

Fig. 8. The chronology of events of a lion pursuing a gazelle.

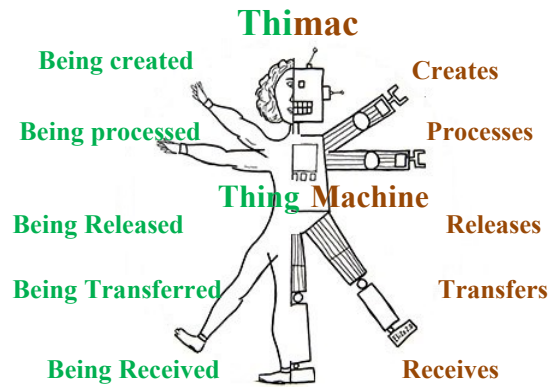

Fig. 9. The Thimac (The figure is adapted from https://engineering.mit.edu/engage/ask-an-engineer/when-will-ai-be-smart-enough-to-outsmart-people/)

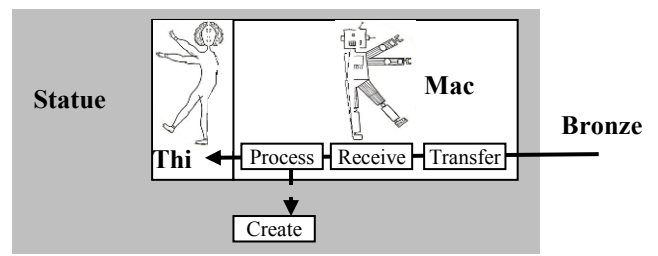

Fig. 10. The *thi* of the bronze statue is formed by its *mac*.

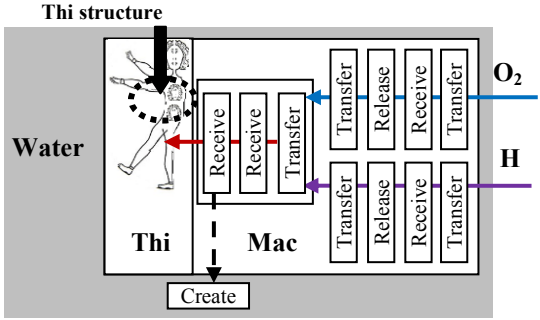

Fig. 11. The *thi* of water is formed by its *mac*.

## IV. THIMACS EXAMPLE

[43] describes the UML as including structural things and behavioral things. Structural things are the "nouns" of UML models and are the mostly stable (static) part of a model (e.g., class, use case, and component). Relationships (dependencies, association, generalization, and realization) are the way in which to connect things. Class diagrams are the central notation for structural aspects [38]. The class connection to behavioral aspects is based on its methods. The UML's modeling capabilities beyond those of classes and their relationships are expanded using composite structure diagrams aimed to model the internal structures of classes [44].

According to Ventura [26], one of the most fundamental qualities of object-oriented software is its contemporary nature—objects typically contain data that reflect the most updated information. Object-oriented software applications handle historical information by listing interaction objects. The "historical" object itself does not contain the ***history*** of the activity that it documents; rather, it contains only the final outcome.

Some cases exist in which it is worthwhile to keep and manage the history of an object, not just its state, at a specific point in time. In the UML, the behavior of an object is the relationship among the sequences of its messages. The communication history is a sequence containing messages and time stamps that mark time progress [38].

Suppose that we would like to develop an application whose purpose is to manage the price list of a furniture manufacturer's product tree. The application's object model is shown in Fig. 2 [26]. We want to use said object model to display the changes in a certain product price during the past year in the form of, for instance, a chart or a graph. It is possible to create a list of objects that provide us with enough information for constructing the desired graph [26]. "We still can't be fully satisfied [since]…management, in the sense of synchronizing between those models (after all, they surely overlap), could become a tedious job. It seems that we should look further and find a way to build object models according to a different approach" [26]. The solution in [26] (see Fig. 3) is a philosophy-based pattern related to the problem of the *essence of identity*, which has two so called *perdurantism* and *endurantism* approaches.

In the TM approach, Fig. 14 shows the furniture *thi* (1). The thing side of the furniture *thimac* is represented as a structure of subthings: the table (2), chair (2), and price (3). This "template" of the thing furniture reflects the "empty shell" that encapsulates the furniture body (i.e., objects). This is roughly similar to the composite structure diagrams [44], in which both cases (TM and composite structure diagrams) are in response to all/parts of the structure. Note that the date is not included because it is an aspect of an event in a TM.

Fig. 15 shows the furniture thimac (it roughly corresponds to the UML class diagram including methods). It receives the data (1) and price (2) of each piece of furniture to create the furniture "object." For example, the create stage (writing the table's serial number and price) is implemented as the informal expression:

*Furniture.Mac.transfer.Receives.(integer).release.transfer."writing table".transfer.receive* (the flow path from input to writing table indicated by the circles 1→3→4→5, and also, in a similar way, the flow path 2→6→7→8.

In conclusion, *furniture* as an entity in the world is a thing and a machine. The thing side has the structure of data and price, and the machine side constructs the thing by receiving data and patches them into writing tables, dining tables, and chairs.

Fig. 16 (after simplifying the thimac diagram of Fig. ?) shows three possible events:
- Event 1 ($E_1$): Instantiation of a dining table object with price
- Event B ($E_B$): Instantiation of a writing table object with price
- Event C ($E_C$): Instantiation of a chair object with price

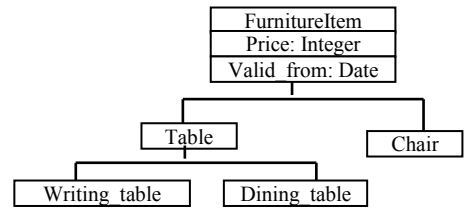

Fig. 12. Example object model (simplified from [26]).

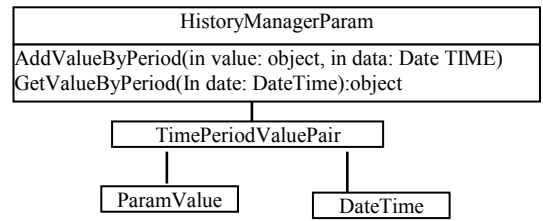

Fig. 13. Part of the proposed solution of the object history management problem (simplified from [26]).

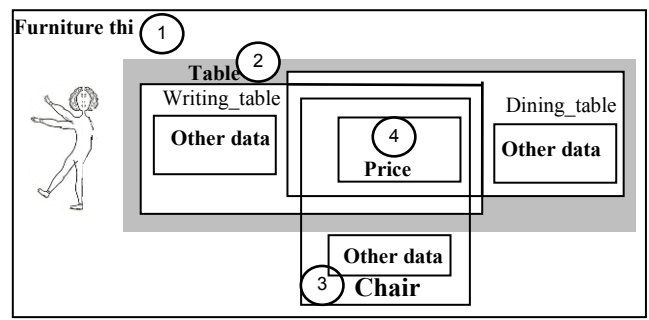

Fig. 14. The structure of the furniture thing.

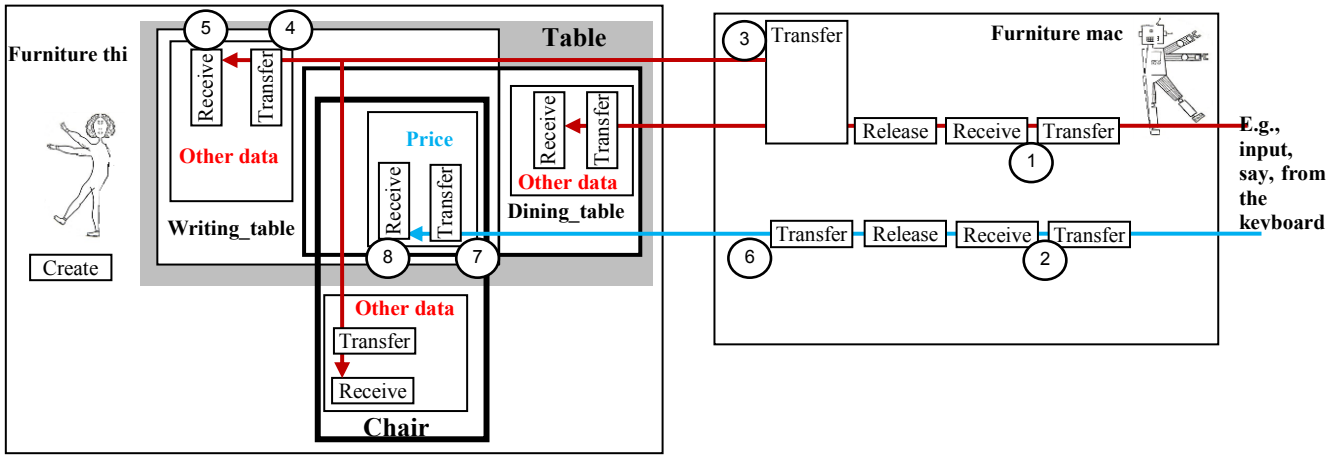

Fig. 15. the Thimac of furniture.

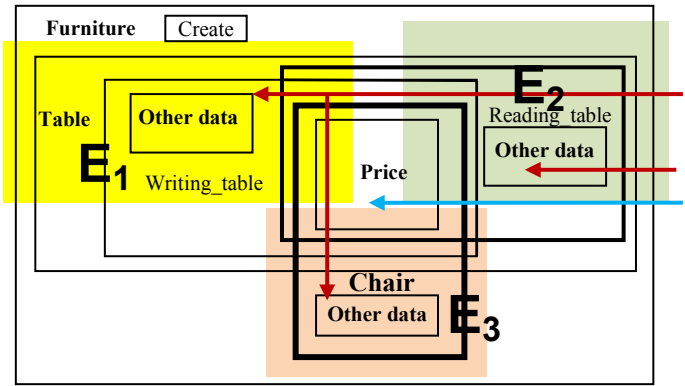

Fig. 16. The events of the furniture thing.

Fig. 17 shows the chronology of these events. Accordingly, the problem of incorporating time into the furniture system becomes an issue of building a database for the furniture events, much like building and analyzing a logging file. The implementation of this history-based database begins with the specification of the system's behavior (Fig. 17).

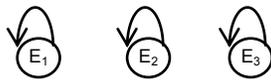

Fig. 17. The chronology of events of the furniture thing.

## V. STRUCTURE 1

Structure may refer to how things, machines, and thimacs fit together to form "larger" things, machines, and thimacs, respectively. Thus, we have thing structure, machine structure, and thimacs structure. A TM model is supposed to be able to make up part of the reality being modeled as seen by the software engineer. In our analysis of structure, we are not concerned with the relationship between a TM model and this reality but instead focus on what is a structure in the context of a given TM description.

Take, for example, a mathematical structure, such as a geometric structure where space comprises a set of points with certain mathematical objects defined on it; these objects give the structure of the space [45]. In a Euclidean plane, we describe the various locations in the plane in terms of a coordinate system, thus associating numerical values with the points in the plane. The relationship between objects and their coordination-dependent descriptions brings structure to these objects.

Accordingly, in a TM, we have to assume that "elementary" thimacs exist. In a given TM model, these elementary things reflect a design decision. For example, we take the thimac *string* (See Fig. 18) as an elementary thimac even though we can consider it to be a thimac that is constructed from *characters*. The elementary thimacs are created without processing any other thimac. The simplest TM structure is the "rootless" machines of these elementary thimacs.

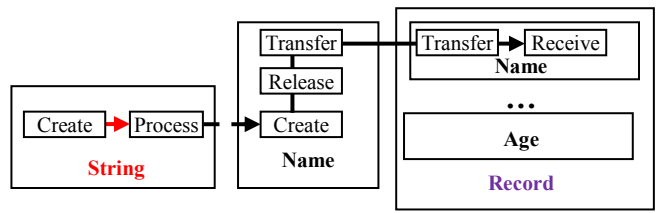

Fig. 18. *String* as an example of an elementary thimac.

Going back to our previous analysis of *t* as a thing in a given system *S*, every *t* has a "region" in *S*. First, we take *S* as the grand TM machine—that is, the totality of the TM diagram that models the part of the reality of concern. Every type of *t* (e.g., order, invoice) has its own *S(t)* that is a subdiagram of *S*. *S(t)* is the subdiagram where *t* (including its subthings) flows.

In fact, every *t* has a place in a stage in a machine in *S*. Thus, as the first step in studying structure, we consider the relationships of *t*s in their regions in the machines. From a structuralism point of view, thimacs must be understood by way of their places in the broader, overarching grand thimac structure.

**Example** (composite structure modeling)**:** According to Kumar and Jasperneite [46], the structural modeling elements of a UML communication system profile supports system, block, process, and package agents. A system's structure is specified using classes and composite structures. The composite structure specifies the communication paths by means of connectors or channels, and it interfaces for signals and remote procedure calls. Fig. 19 provides an example of the usage of a UML communication system profile. Agents modeled with active classes that execute a behavior after initiations are called processes.

The TM diagram produces a more elegant structure without introducing any additional notions as shown in Fig. 20. To emphasize the composite structure of the system, the figure can be simplified by deleting the transfer, receive, and release stages under the assumption that the arrow and its direction is sufficient for indicating the type of input/output flow (see Fig. 21).

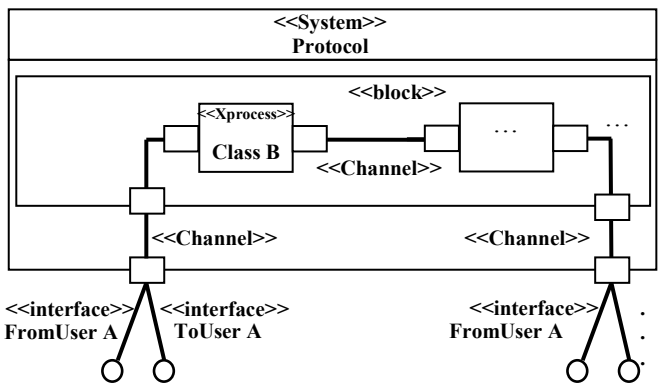

Fig. 19. Composite structure diagram of the example Protocol using the UML communication system profile (partially redrawn from [46]).

## VI. STRUCTURE 2

An important approach to quality assurance for the conceptual models is the ontological foundations of the core concepts. A main success factor of a conceptual modeling language is its ability to provide a set of constructs that enable users to express domain concepts in an unambiguous manner. UML class diagrams and BPMN seem to be the best choices for conceptual modeling [47].

Birta and Arbez [48] introduce the simulation of a conceptual modeling framework that is formulated from "entities that interact over the course of the observation interval by reacting to, and giving rise to, the occurrence of events which are the second important constituent of the framework." They introduce the framework with an example titled Kojo's Kitchen. Kojo's Kitchen serves two types of products: sandwiches and sushi. Two employees currently work at the counter throughout the day. Both of them serve sandwiches and sushi to customers.

Birta and Arbez [48] construct a structural view to identify the entity categories. The structural view consists of a structural diagram followed by a description of each entity category as shown in Fig. 22.

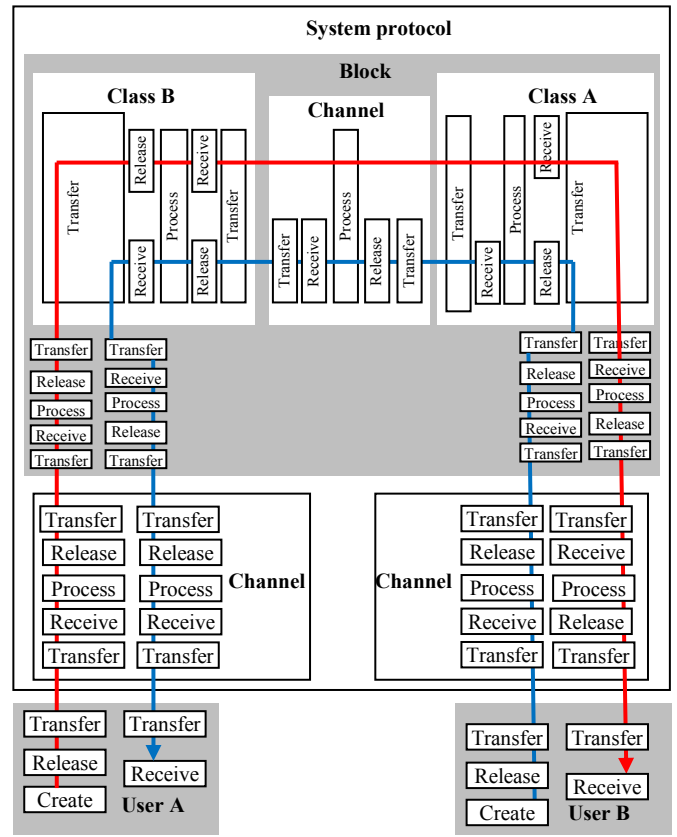

Fig. 20. The TM model of the example protocol.

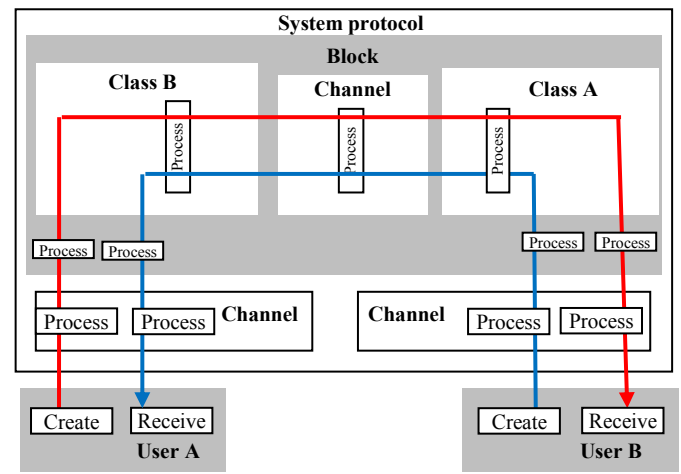

Fig. 21. Simplification of the TM model of the example protocol.

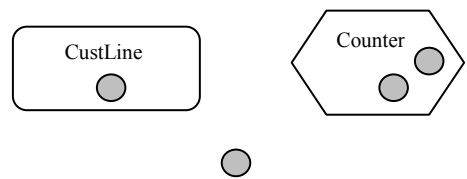

Fig. 22. Structural diagram (redrawn from [48]).

Without the loss of generality, we assume two counters. The Kojo's Kitchen model in the TM is shown in Fig. 23. It has three large submachines: the queue (circle 1) and two counters, A and B (2 and 4, respectively). A new customer arrives (4) at Kojo's Kitchen to enter the queue (5). The action of joining the queue Q involves the following: (The circular queue stored in Q(0:n - 1); rear points to the last item and front is one position counterclockwise from the first item in Q.)

1. The Rear (6) is released (7) to be incremented (8), and the customer with the assigned queue number joins the queue (9). The white box (with red outlines) of the queue denotes some function (routine) that accommodates the customer in the queue using Rear. Additionally, if the Rear after being incremented makes Q full) (10), then the entrance is blocked (11) to prevent additional customers from coming.
2. If either counter A or counter B is available (not busy) (12 and 13), then this would start processing Front (14) to release a customer in the Front of the queue, assuming that the queue is not empty (15 and 16). The vertical thick lines at circles 15 and 16 represent the availability of a counter AND that the queue is not empty. In this case, processing Front (14) triggers the decrementing of the Front (17). If the Q is full (18) (before the decrementing phase), then the blocked entrance is lifted (19).
3. Receiving a customer at counter A blocks this counter (20) and changes its state to busy (21). The customer is processed (22) and released to the outside (23). Releasing the customer (24) frees up and unblocks the counter (25).
4. Counter B has a similar procedure as step 3.

Fig. 23 can be simplified by removing the release, transfer, and receive stages under the assumption that the directions of the arrows are sufficient for indicating the direction of the flow as shown in Fig. 24.

In general, the structure and the behavior of the TM model interweave together such that it is difficult to separate them. As we saw in the furniture example discussed previously, even price and other data (the *thi* side of the furniture in Fig. 15) involve the stages of transfer and receive.

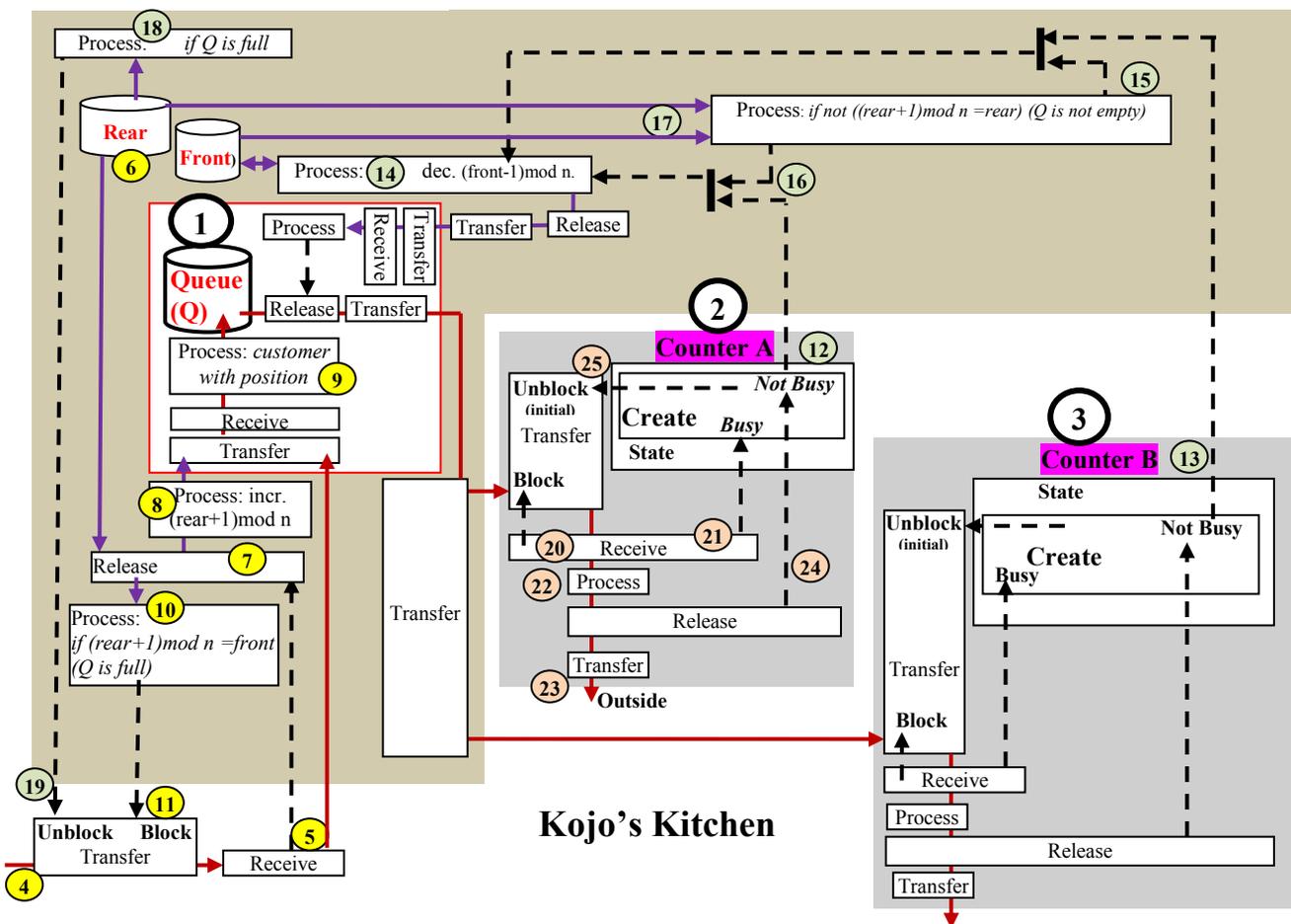

Fig. 23. The TM model of Kojo's Kitchen.

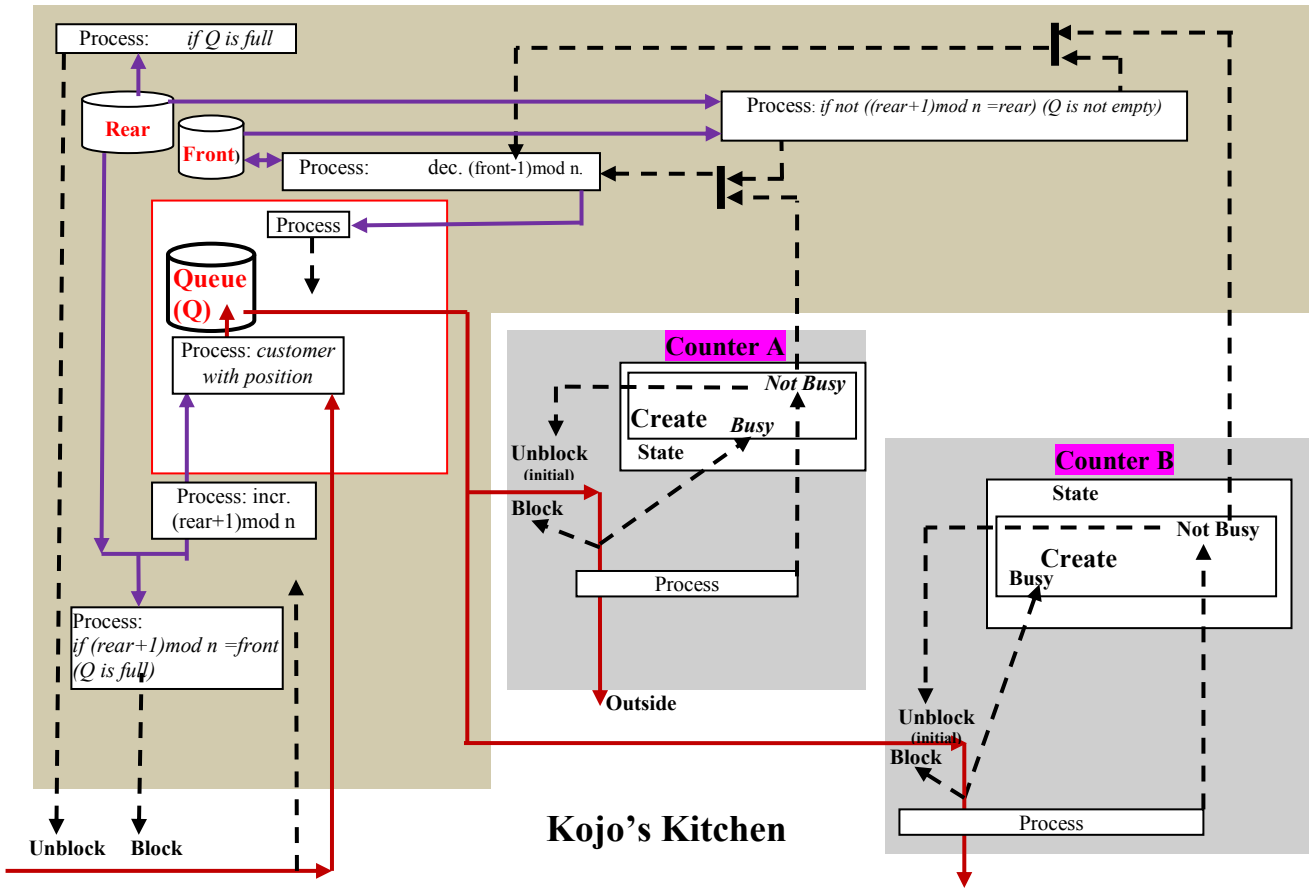

Fig. 24. Simplification of the TM model of Kojo's Kitchen.

VII. CONCLUSION

In this paper, we propose a philosophical approach to the TM model to create a modeling tool that can be used in the software system requirements analysis and design process. By exploring the notion of structure, we find a new approach for the TM model in which we establish a philosophical foundation in software engineering that is worth further research.